\documentclass[twocolumn,aps,superscriptaddress,showpacs,nofootinbib,floatfix]{revtex4}
\usepackage{epsfig,bm,feynmf}
\usepackage{graphics}
\usepackage{amsmath}
\usepackage[normalem]{ulem}  
\usepackage[dvips]{color} 

\renewcommand\sout{\bgroup \color{red} \ULdepth=-.5ex \ULset}

\begin{document}


\title{Light (anti-)nuclei production and flow in relativistic heavy-ion collisions}


\author{Lilin Zhu}\email{zhulilin@scu.edu.cn}
\affiliation{Department of Physics, Sichuan University, Chengdu, China}
\author{Che Ming Ko}\email{ko@comp.tamu.edu}
\affiliation{Cyclotron Institute and Department of Physics and Astronomy, Texas A$\&$M University, College Station, TX 77843, USA}
\author{Xuejiao Yin}\email{yinxuejiao@stu.scu.edu.cn}
\affiliation{Department of Physics, Sichuan University, Chengdu, China}


\begin{abstract}
Using the coalescence model based on the phase-space distributions of protons, neutrons, Lambdas and their antiparticles from a multiphase transport (AMPT) model, we study the production of deuteron, triton, helium 3, hypertriton, hyperhelium 3 and their antinuclei in Pb+Pb collisions at $\sqrt{s_{NN}}=2.76$ TeV. The resulting transverse momentum spectra, elliptic flows and coalescence parameters for these nuclei are presented and compared with available experimental data. We also show the constituent number scaled elliptic flows of these nuclei and discuss their implications.
\end{abstract}

\pacs{25.75.Nq, 25.75.Ld}
\keywords{}
\maketitle

\section{introduction}

Recently, light nuclei production has been studied at the Large Hadron Collider (LHC) by the ALICE Collaboration~\cite{Adam:2015vda,Adam:2015yta}. Similar to the experiments carried out earlier by the STAR Collaboration at the Relativistic Heavy Ion Collider (RHIC)~\cite{Adler:2001uy,Abelev:2009ae,Abelev:2010rv,Agakishiev:2011ib}, the motivation for such studies are twofolds. One is to search for nuclei that do not exist in nature in order to study if nuclei and antinuclei have same properties and to discover the stability of multistrange hypernuclei and antinuclei. The other is to use light nuclei to study the space-time structure of the emission source in relativistic heavy ion collisions since they are likely produced at kinetic freeze out via nucleon coalescence, complimenting the method using the HBT interferometry~\cite{Bertsch:1988db,Pratt:1990zq} of particles emitted at freeze out. The use of the coalescence model for studying light nuclei production has a long history with applications in heavy-ion collisions at both intermediate~\cite{Gyulassy:1982pe,Aichelin:1987ti,Koch90,Indra00} and high energies~\cite{Mattiello:1996gq,Nagle:1996vp} as well as at relativistic energies~\cite{Chen:2012us,Chen:2013oba,Zhang:2009ba,Sun:2015jta,She:2015bta}. In most applications of the coalescence model, the energy spectra for clusters are simply given by the product of the spectra of their constituent nucleons multiplied by an empirical coalescence parameter.  In more sophisticated coalescence model, the coalescence parameter is computed from the overlap of the nuclear Wigner phase-space density with the nucleon phase-space distributions at freeze out. For example, in Refs.~\cite{Chen:2003qj,Chen:2003ava}, using the phase-space distribution from an isospin-dependent transport model for heavy ion collisions at intermediate energies with radiative beams, production of light nuclei such as deuteron, triton, helium 3 and alpha has been studied in the coalescence model. It was found in this study that the yield of light nuclei is sensitive to the density dependence of nuclear symmetry energy.  Also, the study of deuteron production in heavy ion collisions at RHIC was studied in Ref.~\cite{Oh:2009gx} based on a multiphase transport (AMPT) model.  Both the coalescence model based on the phase-space distributions of protons and neutrons at freeze out and a dynamic model that includes deuteron production and annihilation via $NN\leftrightarrow\pi d$ in the hadronic stage of AMPT have been used. It was found that the final deuteron yield and elliptic flow from these two approaches are similar, providing thus a consistent check on the applicability of the coalescence model to deuteron production in heavy ion collisions. In the present study, we generalize the study of Ref.~\cite{Oh:2009gx} to include the production of not only deuteron but also triton, helium 3, hypertriton, and hyperhelium 3 as well as their antinuclei from the coalescence model using the phase-space distribution of protons, neutrons, Lambdas and their antiprticles at freeze out from the AMPT model in both its default and string melting versions. We specifically study the transverse momentum spectra and elliptic flows of these nuclei for Pb+Pb collisions at $\sqrt{s_{NN}}=2.76$ TeV as studied in the experiments by the ALICE Collaboration. We also determine the coalescence parameters from the transverse momentum spectra of these nuclei.

The paper is organized as follows. In Section~\ref{ampt}, we briefly review the AMPT model used for the present study. The coalescence model is then described in Section~III. In Section~\ref{results}, we show results from our study on the transverse momentum spectra, elliptic flows, and coalescence parameters for deuteron, triton, helium 3, hypertriton, and hyperhelium 3 as well as their antinulcei. Finally, a summary is given in Section~\ref{summary}.

\section{the AMPT model}\label{ampt}

To obtain the phase-space distributions of protons, neutrons, and Lambdas as well as their antiparticles, we use the AMPT model that has been extensively utilized for studying heavy ion collisions at relativistic energies. The AMPT model is a hybrid model~\cite{Lin:2004en} with the initial particle distributions generated by the heavy ion jet interaction generator (HIJING) model~\cite{Wang:1991hta}. In the default version, the jet quenching in the HIJING model is replaced in the AMPT model by explicitly taking into account the scattering of mini-jet partons via the Zhang's parton cascade (ZPC) model~\cite{Zhang:1997ej}. These partons are recombined with their parent strings after the scattering, which are then converted to hadrons using the Lund string fragmentation model. In the version of string melting, all hadrons produced from the string fragmentation in the HIJING model are converted to their valence quarks and antiquarks, whose evolution in time and space is modeled by the ZPC model. After the end of their scatterings, quarks and antiquarks are converted to hadrons via a spatial coalescence model. In both versions of the AMPT model, the scatterings among hadrons are described by a relativistic transport (ART) model~\cite{Li:1995pra}.  In the present study, we adopt the version Ampt-v1.25t7-v2.25t7~\cite{ampt} with the default Lund string fragmentation parameters $a=0.5$ and $b=0.9$ GeV$^{-2}$ in the HIJING model, the QCD coupling constant $\alpha_s=0.33$, and the screening mass $\mu=3.2$ fm$^{-1}$ to obtain a parton scattering cross section of 1.5 mb in ZPC. These parameters were shown in Ref.~\cite{Xu:2011fi} to give a better description of both the charged particle multiplicity density and elliptic flow measured in heavy ion collisions at the LHC than their values used in Ref.~\cite{Lin:2004en} for heavy ion collisions at RHIC.

\section{the coalescence model}\label{coal}

For light nuclei production in heavy ion collisions, both the statistical model~\cite{Andronic:2010qu,Cleymans:2011pe}, which assumes that light nuclei are in both thermal and chemical equilibrium with all other particles in the produced hot dense matter, and the coalescence model have been used. In the present study, we use the coalescence model to study light nuclei production based on the phase-space distributions of protons, neutrons, and Lambdas as well as their antiparticles from the AMPT model described in the previous section.

The coalescence model for nuclei production in heavy ion collisions is based on the sudden approximation of projecting out their wave functions from the nucleon wave functions at freeze out.  As shown in Ref.~\cite{Mattiello:1996gq}, the number of nucleus of atomic number $A$ produced in a heavy ion collision is then given by the overlap of the Wigner function $f_A({\bf x}_1^\prime, ... ,{\bf x}_A^\prime; {\bf p}_1^\prime, ... ,{\bf p}_A^\prime)$ of the produced nucleus with the nucleon phase space distribution function $f_N({\bf x},{\bf p})$ at freeze out, that is
\begin{eqnarray}
\label{coal}
&&\frac{dN_A}{d^3 {\mathbf P}_A}= g_A\int \Pi_{i=1}^Ad^3{\bf x}_i d^3{\bf p}_i
f_N({\bf x}_i, {\bf p}_i)\nonumber\\
&&\times f_A({\bf x}_1^\prime, ... ,{\bf x}_A^\prime; {\bf p}_1^\prime, ... ,{\bf p}_A^\prime)\delta^{(3)}\left({\bf P}_A-\sum_{i=1}^A{\bf p}_i\right),\nonumber\\
\end{eqnarray}
where ${\bf x}_i^\prime$ and ${\bf p}_i^\prime$ are the nucleon coordinates ${\bf x}_i$ and momenta  ${\bf p}_i$ in the center of mass frame of the nucleus, and $g_A=(2J_A+1)/2^A$ is the statistical factor for $A$ nucleons of spin $1/2$ to form a nucleus of angular momentum $J_A$.

For the light nuclei we are considering in this study, such as deuteron, triton, helium 3 and their antinuclei as well hypertriton, hyerhelium 3 and their anti-hypernuclei, their wave functions can be approximately given by those of the ground state of a harmonic oscillator. In this case, the Wigner function for a nucleus consisting of two constituent particles is~\cite{Song:2012cd}
\begin{eqnarray}
f_2(\boldsymbol\rho,{\bf p}_\rho)=8g_2\exp\left[-\frac{\boldsymbol\rho^2}{\sigma_\rho^2}-{\bf p}_\rho^2\sigma_\rho^2\right],
\label{two}
\end{eqnarray}
where
\begin{eqnarray}\label{rel}
\boldsymbol\rho=\frac{1}{\sqrt{2}}({\bf x}_1^\prime-{\bf x}_2^\prime),\quad{\bf p}_\rho=\sqrt{2}~\frac{m_2{\bf p}_1^\prime-m_1{\bf p}_2^\prime}{m_1+m_2},\nonumber\\
\end{eqnarray}
with $m_i$, ${\bf x}_i^\prime$ and ${\bf p}_i^\prime$ being the mass, position and momentum of particle $i$, respectively.

Similarly, the Wigner function for a nucleus consisting of three constituent particles is~\cite{Song:2012cd}
\begin{eqnarray}
&&f_3(\boldsymbol\rho,\boldsymbol\lambda,{\bf p}_\rho,{\bf p}_\lambda)\nonumber\\
&&=8^2g_3\exp\left[-\frac{\boldsymbol\rho^2}{\sigma_\rho^2}-\frac{\boldsymbol\lambda^2}{\sigma_\lambda^2}-{\bf p}_\rho^2\sigma_\rho^2-{\bf p}_\lambda^2\sigma_\lambda^2\right],
\label{three}
\end{eqnarray}
where $\boldsymbol\rho$ and ${\bf p}_\rho$ are similarly defined as in Eq.(\ref{rel}), and
\begin{eqnarray}
{\boldsymbol\lambda}&=&\sqrt{\frac{2}{3}}\left(\frac{m_1{\bf x}_1^\prime+m_2{\bf x}_2^\prime}{m_1+m_2}-{\bf x}_3^\prime\right),\nonumber\\
{\bf p}_\lambda&=&\sqrt{\frac{3}{2}}~\frac{m_3({\bf p}_1^\prime+{\bf p}_2^\prime)-(m_1+m_2){\bf p}_3^\prime}{m_1+m_2+m_3}.
\end{eqnarray}

The width parameter $\sigma_\rho$ in Eq.(\ref{two}) is related to the root-mean-square radius of the nucleus of two constituent particles via~\cite{Song:2012cd}
\begin{eqnarray}
\langle r_2^2 \rangle=\frac{3}{2}\frac{m_1^2+m_2^2}{(m_1+m_2)^2}\sigma_\rho^2=\frac{3}{4}\frac{m_1^2+m_2^2}{\omega  m_1m_2(m_1+m_2)},
\end{eqnarray}
where the second line follows if we use the relation $\sigma_\rho=1/\sqrt{\mu_1\omega}$ in terms of the oscillator frequency $\omega$ in the harmonic wave function and the reduced mass $\mu_1=2(1/m_1+1/m_2)^{-1}$.

For the width parameter $\sigma_\lambda$ in Eq.(\ref{three}), it is related to the oscillator frequency by $(\mu_2 \omega)^{-1/2}$, with $\mu_2=(3/2)[1/(m_1+m_2)+1/m_3]^{-1}$. Its value and also that of $\sigma_\rho=1/\sqrt{\mu_1\omega}$ are then determined from the oscillator constant via the root-mean-square radius of the nucleus of three constituent particles, that is~\cite{Song:2012cd}
\begin{eqnarray}
&&\langle r_3^2 \rangle\nonumber\\
&&=\frac{1}{2}\frac{m_1^2(m_2+m_3)+m_2^2(m_3+m_1)+m_3^2(m_1+m_2)}{\omega(m_1+m_2+m_3)m_1m_2m_3}.\nonumber\\
\end{eqnarray}

\section{results}\label{results}

In the present Section, we show results on the transverse momentum spectra, elliptic flows, and coalescence parameters for deuteron, triton, helium 3, hypertriton, and hyperhelium 3 as well as their antinuclei. They are calculated from the coalescence model using the phase-space distributions of proton, neutron, Lambda and their antiparticles obtained from the AMPT model. For deuteron, triton, and helium 3, their statistical factors and the values of the width parameters in their Wigner functions are shown in Table \ref{tab} together with the empirical values of their radii and the resulting oscillator constants.  Same parameters are used for their antinuclei.  For hypertriton and hypehelium 3 as well as their antinuclei, they are assumed to have same properties as corresponding nuclei of three nucleons, and the width parameters in their Wigner functions are thus taken to be the same as those for triton and helium 3, respectively.

\begin{table}[h]
\caption{{\protect\small Statistical factor ($g$), radius ($R$), oscillator frequency ($\omega$), and width parameter ($\sigma_\rho$, $\sigma_\lambda$) for deuteron ($d$), triton($t$), and helium 3 ($^3He$).}} \label{tab}
\begin{tabular}{c|cccccc}
\hline\hline
Nucleus & $g$ & R (fm) & $\omega$ (sec$^{-1})$ & $\sigma_\rho, \sigma_\lambda$ (fm) \\
\hline
$d$ & 3/4 & 1.96 & 0.2077 & 2.263 \\
$t$  & 1/4 & 1.61 & 0.4104 & 1.61 \\
$^3He$ & 1/4 & 1.74 & 0.3514 & 1.74 \\
\hline\hline
\end{tabular}
\end{table}

Since the hot dense matter produced in the midrapidity of Pb+Pb collisions at $\sqrt{s_{NN}}=2.76$ TeV is essentially baryon free and has zero isospin, the distributions of protons and neutrons as well as their antiparticles are similar. Therefore, deuterons and antideuterons also have similar distributions. The same is true for the distributions of tritons, helium 3, and their antiparticles as well as for the distributions of hypertritons, hyperhelium 3, and their antiparticles. In the following, we thus show results that are obtained by averaging over these similar distributions, i.e., $(p+n+\bar p+\bar n)/4$, ($\Lambda+\bar\Lambda)/2$, $(d+\bar d)/2$, $(t+He+\bar t+^3\overline{He})/4$, and $(^3_\Lambda H+^3_\Lambda He+^3_{\bar\Lambda}\bar H+^3_{\bar\Lambda}\overline{He})/4$, and they are called nucleon-like, Lambda-like, deuteron-like, triton-like, and hyper-triton-like, respectively.

\subsection{Transverse momentum spectra}

\begin{figure}[h]
\centerline{
\includegraphics[width=8.5cm]{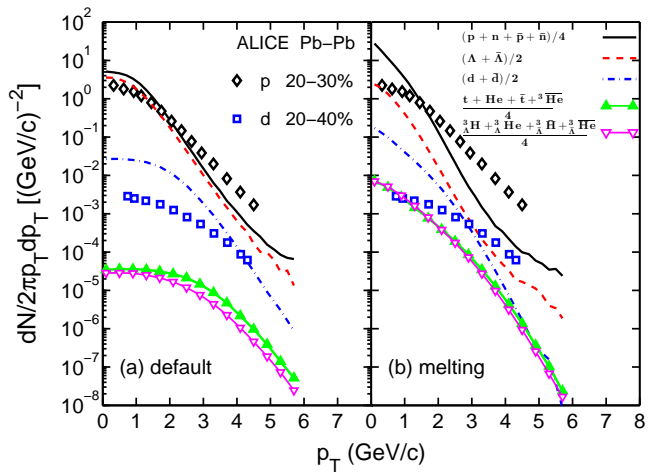}}
\caption{(Color online) Transverse momentum spectra of $(p+n+\bar p+\bar n)/4$ (solid line), $(\Lambda+\bar\Lambda)/2$ (dashed line), $(d+\bar d)/2$ (dash-dotted line), $(t+He+\bar t+^3\overline{He})/4$ (filled triangles), and $(^3_\Lambda H+^3_\Lambda He+^3_{\bar\Lambda}\bar H+^3_{\bar\Lambda}\overline{He})/4$ (open triangles) at midrapidity $|y|\le 0.5$ from the default (left panel) and the string melting (right panel) AMPT model for Pb+Pb collisions at $\sqrt{s_{NN}}=2.76$ TeV and impact parameter $b=8$ fm. Data for protons (open diamonds) and deuterons (open squares) are from the ALICE Collaboration~\cite{Adam:2015vda,Adam:2015yta}.}
\label{pt}
\end{figure}

In Fig.~\ref{pt}, we show the transverse momentum spectra of $(p+n+\bar p+\bar n)/4$ (solid line), $(\Lambda+\bar\Lambda)/2$ (dashed line), $(d+\bar d)/2$ (dash-dotted line), $(t+He+\bar t+^3\overline{He})/4$ (filled triangles), and $(^3_\Lambda H+^3_\Lambda He+^3_{\bar\Lambda}\bar H+^3_{\bar\Lambda}\overline{He})/4$ (open triangles) at midrapidity from the default (left panel) and the string melting (right panel) AMPT model for Pb+Pb collisions at $\sqrt{s_{NN}}=2.76$ TeV and impact parameter $b=8$ fm. Also shown in the figure are the proton (open diamonds) and deuteron (open squares) transverse momentum spectra from the ALICE Collaboration~\cite{Adam:2015vda} for collisions at centralities 20-30\% and 20-40\%, respectively, which are similar to collisions at impact parameter $b=8$ fm used in the AMPT calculations. It is seen that the default AMPT model describes reasonably the experimentally measured proton transverse momentum spectrum but overestimates that of deuteron by about a factor of two.  The string melting AMPT model overestimates, however, both measured proton and deuteron transverse momentum spectra.  The latter is not surprising as it has already been pointed out in Ref.~\cite{Lin:2004en} that baryons are not properly described by AMPT with string melting, giving generally a larger number and a soft transverse momentum spectrum at midrapidity, because the way AMPT treats baryon production via quark coalescence at hadronization. Although some improvements on the problem have been introduced in the version of AMPT code used in the present study by changing the coalescence order between mesons and baryons, this has apparently not solved the problem. Further improvements are thus needed in the AMPT model.

The total number of light nuclei produced in a collision can be obtained from integrating their transverse momentum spectra. For the default AMPT model, the numbers are 24 for the nucleon-like, 16 for the Lambda-like, 0.36 for deuteron-like nuclei, $9.2\times 10^{-4}$ for triton-like nuclei, and $5.8\times 10^{-4}$ for hypertriton-like nuclei, while for the string melting AMPT model, the corresponding numbers are 37, 20, 2.1, $2.5\times 10^{-2}$, and $2.7\times 10^{-2}$, respectively. The penalty in adding a nucleon to form a heavier nucleus is thus about two order of magnitude smaller in both the default and the string melting AMPT model, similar to that found in the experimental data.

\subsection{Elliptic flows}

\begin{figure}[h]
\centerline{
\includegraphics[width=8.5cm]{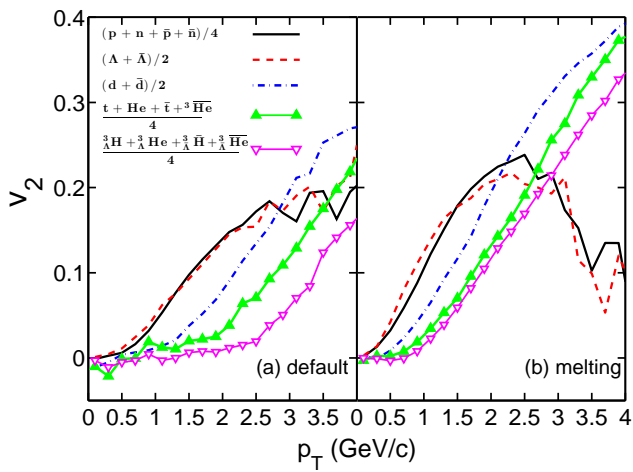}}
\caption{(Color online) Elliptic flow of $(p+n+\bar p+\bar n)/4$ (solid line), $(\Lambda+\bar\Lambda)/2$ (dashed line), $(d+\bar d)/2$ (dash-dotted line), $(t+He+\bar t+^3\overline{He})/4$ (filled triangles), and $(^3_\Lambda H+^3_\Lambda He+^3_{\bar\Lambda}\bar H+^3_{\bar\Lambda}\overline{He})/4$ (open triangles) at midrapidity $|y|\le 0.5$ from the default (left panel) and the string melting (right panel) AMPT model for Pb+Pb collisions at $\sqrt{s_{NN}}=2.76$ TeV and impact parameter $b=8$ fm.}
\label{v2}
\end{figure}

The momentum distribution of nucleus $A$ produced in a heavy ion collision event can be generally written as
\begin{eqnarray}
&&f_A(p_T,\phi,y)=\nonumber\\
&&\frac{N_A(p_T,y)}{2\pi}\left\{1+2\sum_n v_n(p_T,y)\cos[n(\phi-\Psi_n)]\right\},
\end{eqnarray}
where $\phi$ is the azimuthal angle, $\Psi_n$ is the $n$th-order event plane angle, and $N_A(p_T,y)$ and $v_n(p_T,y)$ are the
number of nuclei of transverse momentum $p_T$ and rapidity $y$ and their $n$th-order anisotropic flows, respectively. In the present study, we are only interested in the elliptic flow $v_2$. Also, we neglect the fluctuation of event plane angle $\Psi_2$ by taking $\Psi_2=0$ as our calculations involve the mixing of many events to reduce the statistical fluctuations due to the small number of nucleons and Lambdas in an event.  In this case, the elliptic flow can be simply calculated from
\begin{eqnarray}
v_2(p_T)=\Big\langle\frac{p_x^2-p_y^2}{p_x^2+p_y^2}\Big\rangle,
\end{eqnarray}
where $\langle ...\rangle$ indicates average over all nuclei $A$ in all events.

Figure~\ref{v2} shows the elliptic flow of $(p+n+\bar p+\bar n)/4$ (solid line), $(\Lambda+\bar\Lambda)/2$ (dashed line), $(d+\bar d)/2$ (dash-dotted line), $(t+He+\bar t+^3\overline{He})/4$ (filled triangles), and $(^3_\Lambda H+^3_\Lambda He+^3_{\bar\Lambda}\bar H+^3_{\bar\Lambda}\overline{He})/4$ (open triangles) at midrapidity $|y|\le 0.5$ from the default (left panel) and the string melting (right panel) AMPT model for Pb+Pb collisions at $\sqrt{s_{NN}}=2.76$ TeV and impact parameter $b=8$ fm. These results are obtained from 40,000 AMPT events with the mixing of 50 events in calculating the elliptic flows of light (anti-)nuclei. Both the nucleon-like and the Lambda-like are seen to have similar elliptic flows in both the default and the string melting AMPT model. Also, the heavier are the nuclei, the smaller is their elliptic flow, similar to the mass ordering of elliptic flows seen in the hydrodynamic description of collective flow. Because of the strong partonic scattering, the elliptic flow of nuclei is larger in the string melting version of AMPT than in the default version.

A special feature of the coalescence model is its prediction of an approximate constituent number scaling of particle elliptic flows, which states that the elliptic flow of a composite particle at transverse momentum $p_T$ per constituent is the same as a function of $p_T$ divided by the number of constituents.  For light nuclei considered here, $v_2^A(p_T/A)/A$ is then the same. This scaling would be exact if only constituents of same momentum can coalescence to form a nucleus, corresponding to a width parameter in the Wigner function of the nucleus that is infinitely large.

\begin{figure}[h]
\centerline{
\includegraphics[width=8.5cm]{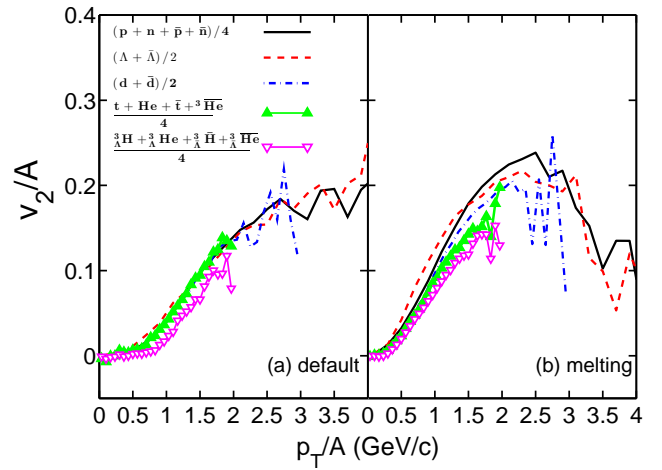}}
\caption{(Color online) Nucleon number scaled elliptic flow of $(p+n+\bar p+\bar n)/4$ (solid line), $(\Lambda+\bar\Lambda)/2$ (dashed line), $(d+\bar d)/2$ (dash-dotted line), $(t+He+\bar t+^3\overline{He})/4$ (filled triangles), and $(^3_\Lambda H+^3_\Lambda He+^3_{\bar\Lambda}\bar H+^3_{\bar\Lambda}\overline{He})/4$ (open triangles) at midrapidity $|y|\le 0.5$ from the default (left panel) and the string melting (right panel) AMPT model for Pb+Pb collisions at $\sqrt{s_{NN}}=2.76$ TeV and impact parameter $b=8$ fm.}
\label{sv2}
\end{figure}

Figure~\ref{sv2} shows the nucleon number scaled elliptic flow of $(p+n+\bar p+\bar n)/4$ (solid line), $(\Lambda+\bar\Lambda)/2$ (dashed line), $(d+\bar d)/2$ (dash-dotted line), $(t+He+\bar t+^3\overline{He})/4$ (filled triangles), and $(^3_\Lambda H+^3_\Lambda He+^3_{\bar\Lambda}\bar H+^3_{\bar\Lambda}\overline{He})/4$ (open triangles) at midrapidity $|y|\le 0.5$ from the default (left panel) and the string melting (right panel) AMPT model for Pb+Pb collisions at $\sqrt{s_{NN}}=2.76$ TeV and impact parameter $b=8$ fm. It indeed shows that the scaled elliptic flows of all light nuclei are similar in the default AMPT model, although there are appreciable deviations in the case of string melting AMPT model, which may again be related to the baryon problem in this model as discussed in the above.

\subsection{Coalescence parameters}

Results from the coalescence model can be characterized by the coalescence parameter $B_A$ defined
via the relation
\begin{eqnarray}
E_A\frac{d^3N_A}{d{\bf p}_A^3}=B_A\left(E_p\frac{d^3N_p}{dp_{\bf p}^3}\right)^A,
\end{eqnarray}
where ${\bf p}_A$ and ${\bf p}_p$ are the momenta of the composite particle and the coalescence constituent, respectively.
Using $d^3{\bf p}/E=dy d^2p_T$, the above equation can be written as
\begin{eqnarray}
\frac{d^3N_A}{dyd^2p_{TA}}=B_A\left(\frac{d^3N_p}{dy_pd^2p_{Tp}}\right)^A.
\end{eqnarray}

\begin{figure}[h]
\centerline{
\includegraphics[width=8.5cm]{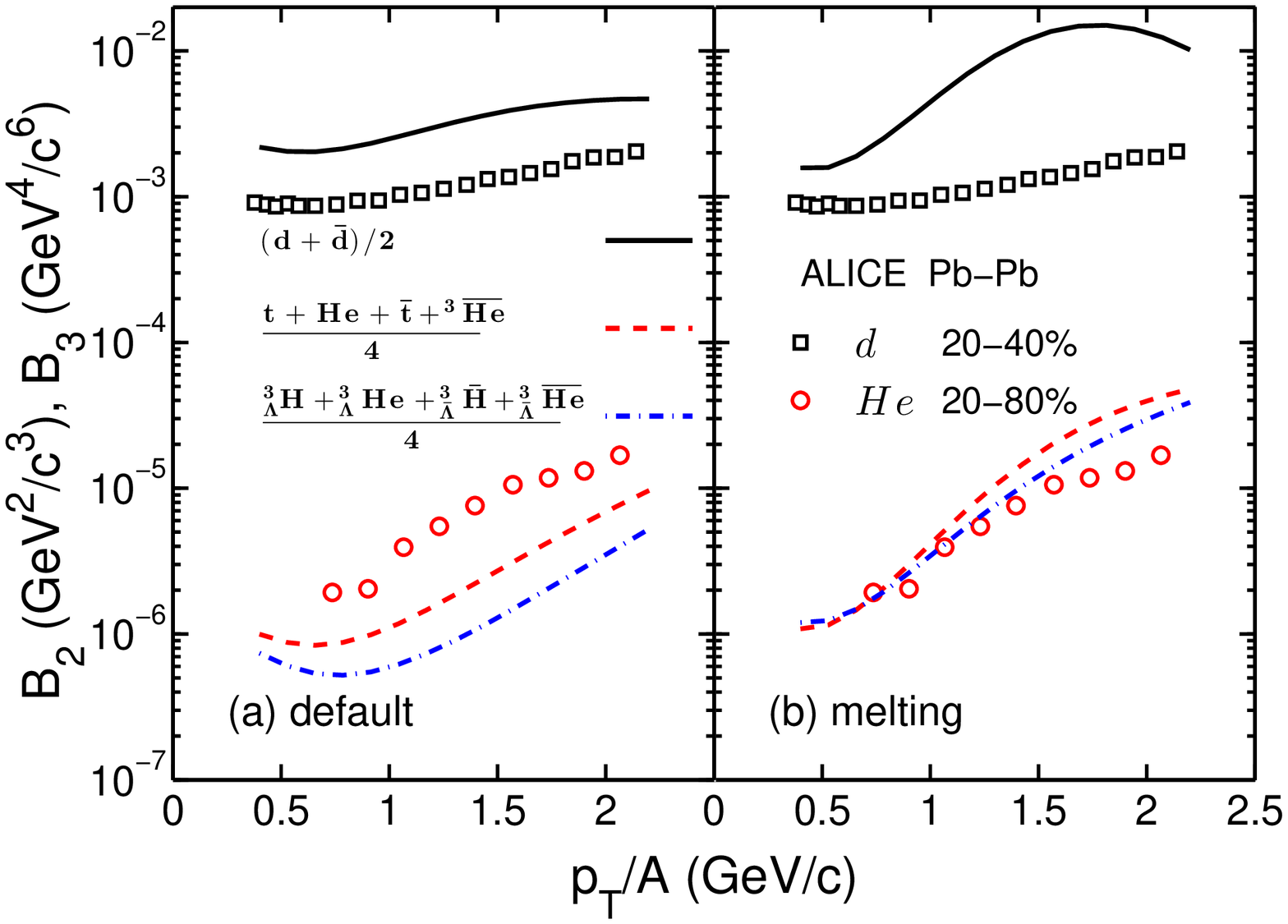}}
\caption{(Color online) Coalescence parameter for $(d+\bar d)/2$ (solid line), $(t+He+\bar t+^3\overline{He})/4$ (dashed line), and $(^3_\Lambda H+^3_\Lambda He+^3_{\bar\Lambda}\bar H+^3_{\bar\Lambda}\overline{He}$ (dash-dotted line) at midrapidity $|y|\le 0.5$ from the default (left panel) and the string melting (right panel) AMPT model for Pb+Pb collisions at $\sqrt{s_{NN}}=2.76$ TeV and impact parameter $b=8$ fm. Data for $B_2$  (open squares) and $B_3$ (open circles) are from the ALICE Collaboration~\cite{Adam:2015vda,Adam:2015yta}.}
\label{ba}
\end{figure}

In Fig.~\ref{ba}, we show the coalescence parameter for $(d+\bar d)/2$ (solid line), $(t+He+\bar t+^3\overline{He})/4$ (dashed line), and $(^3_\Lambda H+^3_\Lambda He+^3_{\bar\Lambda}\bar H+^3_{\bar\Lambda}\overline{He}$ (dash-dotted line) at midrapidity $|y|\le 0.5$ from the default (left panel) and the string melting (right panel) AMPT model for Pb+Pb collisions at $\sqrt{s_{NN}}=2.76$ TeV and impact parameter $b=8$ fm. It is seen that the $B_2$ for deuteron-like nuclei as well as the $B_3$ for both triton-like and hyper triton-like nuclei increase with increasing transverse momentum in both the default and string melting AMPT model, similar to that extracted from the experimental data from the ALICE Collaboration~\cite{Adam:2015vda,Adam:2015yta} shown by open squares for $B_2$ and open circles for $B_3$. Their values at low $p_T$ are a few times 10$^{-3}$ GeV$^2/c^3$ for $B_2$ and about 10$^{-6}$ GeV$^4/c^6$ for the $B_3$ with that for triton-like nuclei slightly larger  than that for hypertriton-like nuclei. Compared to the experimentally extracted values from the ALICE Collaboration, the default AMPT gives a $B_2$ that is about a factor of two larger, similar to that for the transverse momentum spectrum of deuteron-like nuclei, and a $B_3$ that is about a factor of two smaller. For the string melting AMPT model, the obtained $B_2$ is almost an order of magnitude larger than the empirical value, although it gives a $B_3$ that agrees with the empirical one.

\section{summary}\label{summary}

We have studied in the present paper the production of light normal and hypernuclei and their antinuclei in heavy ion collisions at the LHC by using the coalescence model.  With the phase-space distribution of protons, neutrons, and Lambdas as well as their antiparticles at freeze out taken from the AMPT model and taking the Wigner functions of these nuclei to be of Gaussian form with the width parameter fitted from their known radii, we have calculated the transverse momentum spectra and elliptic flows of deuteron-like nuclei that include the deuteron and antideuteron, of triton-like nuclei that include triton and helium 3 and their antinuclei, of hypertriton-like nuclei that include hypertriton and hyperhelium3 and their antinuclei.

For the transverse momentum spectra, we have found that the default version of the AMPT model gives a better description of the experimental data from the ALICE Collaboration for proton and deuteron than the string melting version of the AMPT model, and this has been attributed to the baryon problem in the current string melting version of the AMPT code. From the total yield of these nuclei, we have verified the experimental observation that the yield of light nuclei is reduced by about two orders of magnitude with the addition of a nucleon or Lambda to a nucleus.

For the elliptic flows of these nuclei, they are found to show a mass ordering behavior with the heavy nuclei having a smaller elliptic flow, like that in the hydrodynamical description of heavy ion collisions. This behavior is seen in both the default and the string melting AMPT model. We have further found that the elliptic flow of light nuclei displays an approximate constituent number scaling in that their elliptic flows at transverse momentum $p_T$ per constituent is the same as a function of $p_T$ divided by the number of constituents, particularly in the case of the default AMPT model.

We have further studied the coalescence parameter $B_A$ for light nuclei, which is defined by the ratio of their invariant transverse momentum spectrum to that of their constituents raised to the power corresponding to the number of constituents in the nuclei. Our results based on both the default and string melting AMPT models indicate that the coalescence parameter increases with increasing transverse momentum of the nuclei, similar to that extracted from the experimental data. The values of the coalescence parameters are, however, a factor of two larger for $B_2$ and a factor of two smaller for $B_3$ in the case of the default AMPT model. In the string melting version of the AMPT model, the value of $B_2$ is about an order of magnitude smaller than data but that of $B_3$ agrees with the data.

Although our results are qualitatively comparable to the experimental data from the LHC, they do not give a quantitative description, particularly in the case of the string melting version of the AMPT model due to its problem in treating baryon production during hadronization. Since it is known that a strongly interacting partonic stage exists in relativistic heavy ion collisions, an improved description of baryon production is urgently needed in order to study quantitatively light nuclei production in relativistic heavy ion collisions

Also, the elliptic flows of light nuclei that are produced via the coalescence model are always positive, even though they do show a mass ordering as in the hydrodynamic approach. Since masses of these light nuclei are comparable to that of $J/\psi$, which has been shown to have a negative elliptic flow in the hydrodynamic description of relativistic heavy ion collisions~\cite{Song:2010er}, we expect that the light nuclei studied here would have negative elliptic flows  as well if they are produced statistically in the hydrodynamic model. Therefore, it is of great interest to measure experimentally the elliptic flows of light nuclei to see if they are positive like in the coalescence model or negative as in the hydrodynamic model.

\section*{Acknowledgements}

We thank Lie-Wen Chen, Su Houng Lee, Zie-Wei Lin, Yongseok Oh, and Jun Xu for helpful communications and/or discussions. One of the authors (C.M.K.) is grateful to the Physics Department at Sichuan University for the warm hospitality during his visit when this work was carried out.  This work was supported in part by the NSFC of China under Grant no. 11205106 and the Welch Foundation under Grant No. A-1358.

\end{document}